\documentclass[twocolumn]{aastex63}

\usepackage{lineno}

\usepackage{amsmath}

\usepackage{xspace}

\usepackage{color}

\newcommand{\MESA}{{\texttt{MESA}}\xspace{}}
\newcommand{\COSMIC}{{\texttt{COSMIC}}\xspace{}}
\newcommand{\nmergermesa}{\ensuremath{N_{\mathrm{mrg}}^\mathtt{MESA}}}
\newcommand{\nmergercosmic}{\ensuremath{N_{\mathrm{mrg}}^\mathtt{COSMIC}}}
\newcommand{\Rmesa}{\ensuremath{\mathcal{R}_{\mathrm{mrg}}^\mathtt{MESA}}}
\newcommand{\Rcosmic}{\ensuremath{\mathcal{R}_{\mathrm{mrg}}^\mathtt{COSMIC}}}
\newcommand{\tHubble}{\ensuremath{\tau_{\mathrm{H}}}}
\newcommand{\numberratio}{{number ratio}}
\newcommand{\relativerate}{{relative rate}}
\newcommand{\mathR}{\ensuremath{\mathcal{R}_{\mathrm{rel}}}}
\newcommand{\mathN}{\ensuremath{\mathcal{N}}}
\newcommand{\highlight}{ }

\newcommand{\be}{\begin{enumerate}}
\newcommand{\ee}{\end{enumerate}}

\shorttitle{Binary Black Hole Formation with Detailed Modeling}
\shortauthors{Gallegos-Garcia et al.}

\begin{document}

\title{Binary Black Hole Formation with Detailed Modeling: Stable Mass Transfer Leads to Lower Merger Rates} 
\author[0000-0003-0648-2402]{Monica Gallegos-Garcia}
\affiliation{Department of Physics and Astronomy, Northwestern University, 2145 Sheridan Road, Evanston, IL 60208, USA}
\affiliation{Center for Interdisciplinary Exploration and Research in Astrophysics (CIERA),1800 Sherman, Evanston, IL 60201, USA}

\author[0000-0003-3870-7215]{Christopher P L Berry}
\affiliation{Department of Physics and Astronomy, Northwestern University, 2145 Sheridan Road, Evanston, IL 60208, USA}
\affiliation{Center for Interdisciplinary Exploration and Research in Astrophysics (CIERA),1800 Sherman, Evanston, IL 60201, USA}
\affiliation{SUPA, School of Physics and Astronomy, University of Glasgow, Glasgow G12 8QQ, UK}

\author[0000-0002-0338-8181]{Pablo Marchant}
\affiliation{Institute of Astrophysics, KU Leuven, Celestijnenlaan 200D, B-3001, Leuven, Belgium}

\author[0000-0001-9236-5469]{Vicky Kalogera}
\affiliation{Department of Physics and Astronomy, Northwestern University, 2145 Sheridan Road, Evanston, IL 60208, USA}
\affiliation{Center for Interdisciplinary Exploration and Research in Astrophysics (CIERA),1800 Sherman, Evanston, IL 60201, USA}

\begin{abstract}
Rapid binary population synthesis codes are often used to investigate the evolution of compact-object binaries.
They typically rely on analytical fits of single-star evolutionary tracks and parameterized models for interactive phases of evolution (e.g., mass-transfer on a thermal timescale, determination of dynamical instability, and common envelope) that are crucial to predict the fate of binaries. 
These processes can be more carefully implemented in stellar structure and evolution codes such as \MESA{}. 
To assess the impact of such improvements, we compare binary black hole mergers as predicted in models with the rapid binary population synthesis code \COSMIC{} to models ran with \MESA{} simulations through mass transfer and common-envelope treatment. 
We find that results significantly differ in terms of formation paths, the orbital periods and mass ratios of merging binary black holes, and consequently merger rates. 
While common-envelope evolution is the dominant formation channel in \COSMIC{}, stable mass transfer dominates in our \MESA{} models.
Depending upon the black hole donor mass, and mass-transfer and common-envelope physics, at sub-solar metallicity \COSMIC{} overproduces the number of binary black hole mergers by factors of $2$--$35$ with a significant fraction of them having merger times orders of magnitude shorter than the binary black holes formed when using detailed \MESA{} models. Therefore we find that some binary black hole merger rate predictions from rapid population syntheses of isolated binaries may be overestimated by factors of $\sim\,5$--$500$. We conclude that the interpretation of gravitational-wave observations requires the use of detailed treatment of these interactive binary phases. 

\end{abstract}


\newpage

\section{Introduction}

Binary black hole (BBH) mergers have now been detected through gravitational-wave (GW) observations \citep{Abbott2016}. 
The LIGO Scientific and Virgo Collaboration (LVC) have completed the third observing run, and their current catalog contains over $50$ GW candidates from  compact-object coalescences, where $47$ correspond to BBH mergers \citep{Abbott2019,Abbott2020,Abbott2021}. 
Several independent groups have analyzed the public GW data set \citep{Abbott2021_opendata} and found additional BBH candidates \citep{Venumadhav2019,Zackay2019,Zackay2019_twoBBHmergers,Venumadhav2020,Nitz2020,Nitz2021}. 
In the coming years, the sample of BBH mergers will grow rapidly as the sensitivity of the GW detectors improves \citep{Abbott2020_sensitivity}. 
Once we have collected a sample of hundreds of detections, the challenge will be to accurately interpret and understand these observations.

An outstanding question is how the detected BBHs formed---many channels have been proposed. 
These can broadly be separated into two categories: isolated binary evolution and formation involving dynamical interactions. 
Isolated binary evolution includes formation following a common-envelope (CE) or stable mass-transfer {\highlight(MT)} phase \citep[][]{Paczynski1976,vandenHeuvel1976,Tutukov1993,Belczynski2002,Dominik2012,Stevenson2017,Giacobbo2018,VignaGomez2020,vandenHeuvel2017, neijssel_effect_2019,bavera2021}, or following chemically homogeneous evolution \citep{deMink2016,Marchant2016,Mandel2016,duBuisson2020,Riley2020}, and may include Population III stars \citep{Belczynski2004,Kinugawa2014,Inayoshi2017}. 
Formation of merging BBHs through dynamical interactions includes systems within globular clusters {\highlight \citep{Kulkarni1993,Sigurdsson1993,PortegiesZwart2000,Rodriguez2015,Rodriguez2021, Fragione2018,DiCarlo2019}}, isolated triple {\highlight and quadruple systems \citep{Thompson2011,Antonini2017, Fragione2019, VignaGomez2021},} young stellar clusters \citep{Rastello2020,Banerjee2021,Trani2021}, nuclear star clusters \citep{Antonini2016,ArcaSedda2018,Zhang2019}, and within disks of active galactic nuclei {\highlight \citep{Stone2017,Bartos2017,Fragione2019a,Grobner2020,Kaaz2021}.} 
The overall population likely includes a mix of channels \citep{Zevin2021}, and understanding observations requires detailed modeling of each.

We focus on the isolated binary evolution of BBHs through CE evolution and stable MT \citep[for a review, see][]{Postnov2014,Mapelli2018}. 
To interpret GW observations through any channel requires simulation of large binary populations, and hence the use of population synthesis codes. 
To model stellar and binary evolution efficiently, many population synthesis codes implement single-star evolution formulae based upon \citet{Hurley2000} combined with prescriptions to model binary evolution physics. 
These include {{\texttt{BSE}}\xspace{}} \citep{Hurley2002}, {{\texttt{StarTrack}}\xspace{}} \citep{Belczynski2002,belczynski_compact_2008}, {{\texttt{binary$\textunderscore$c}}\xspace{}}  \citep{Izzard2004,Izzard2006,Izzard2009}, {{\texttt{MOBSE}}\xspace{}} \citep{Giacobbo2018,Giacobbo2018_eddington},  {{\texttt{COMPAS}}\xspace{}}  \citep{Stevenson2017,Barrett2018}, and the \texttt{Compact Object Synthesis and Monte Carlo Investigation Code} \citep[\COSMIC{};][]{Breivik2020}. 
Similarly, {{\texttt{SEBA}}\xspace{}} \citep{PortegiesZwart1996,Nelemans2001} uses the formulae based upon \citet{Eggleton1989}.
Others such as {{\texttt{BPASS}}\xspace{}} \citep{Eldridge2009,Eldridge2017,StanwayEldridge2018}, {{\texttt{SEVN}}\xspace{}} \citep{Spera2015,Spera2019}, {{\texttt{COMBINE}}\xspace{}} \citep{Kruckow2018} and {{\texttt{METISSE}}\xspace{}} \citep{Agrawal2020} use a variety of more detailed stellar evolution models.
Population synthesis codes such as these have been instrumental in advancing our understanding of double compact-object populations for the past few decades. 

Although binary populations synthesis codes have been necessary to statistically study compact-object formation, they also have uncertainties and shortcomings. 
Some uncertainties, such as in stellar winds \citep{Renzo2017,Vink2021}, 
reflect an incomplete  understanding of the relevant physics involved \citep[for a review of massive star winds, see][]{Puls2008}. 
However, other uncertainties, such as in  MT stability and efficiency \citep{olejak2021impact,Garcia2021} or the mass boundary at which CE evolution
terminates \citep{Han1994,DewiTauris2000,Ivanova2011}, may be improved by using detailed modeling of binary evolution.
Understanding these sources of uncertainty is central to accurately interpreting observations.

Recently, detailed simulations of binary evolution have shown that MT is stable over wide ranges in orbital period and mass ratio, avoiding CE \citep{WoodsIvanova2011,ge_adiabatic_2015,Pavlovskii2017,vandenHeuvel2017,Marchant2021}. 
Using the state-of-the-art stellar evolution code \texttt{Modules for Experiments in Stellar Astrophysics} \citep[\MESA{};][]{Paxton2011,Paxton2013,Paxton2015,Paxton2019} some studies have concluded that the occurrence of successful CE leading to BBH merger may be overestimated in population synthesis codes \citep{Klencki2020a,Klencki2021,Marchant2021}. 
\citet{Klencki2021} showed that even with optimistic assumptions for CE, only donors with convective envelopes survive a CE phase. 
In addition, \citet{Klencki2020a} showed that a specific stellar phase (e.g., core helium burning stars, Hertzsprung gap stars) does not directly correlate with a convective envelope, an assumption often made in population synthesis codes. 
\citet{Marchant2021} modeled MT and CE evolution of a massive $30 M_{\odot}$ main-sequence (MS) donor with varying point-mass companions at a wide range of orbital periods. 
They found that stable MT dominates the formation rate of merging BBHs.
Similar results are shown with population synthesis codes using updated CE stability criteria \citep{neijssel_effect_2019,olejak2021impact,Shao2021}.
In an effort to include up-to-date methods of binary evolution into studies of large binary populations, groups have implemented hybrid methods in their studies. 
These methods typically use population synthesis codes combined with detailed simulations of stellar and binary evolution \citep{Nelson2012,Chen2014,Shao2019,bavera2021,RomanGarza2021,Zapartas2021,Shao2021}.
These studies show that including more detailed modeling of binary interactions may reveal details that are missed using simpler prescriptions.

We use a hybrid approach to study the effect of detailed modeling of stellar and binary physics on BBH mergers. 
We address the question: how will the formation of merging BBHs be affected if we use detailed simulations instead of the prescriptions of \COSMIC{} to model the evolution of hydrogen-rich donors with black hole companions (BH--H-rich star binaries)?
Using similar binary evolution treatment as \cite{Marchant2021}, we compare predicted BBH mergers between the rapid population synthesis code \COSMIC{} and \MESA{} models. 
With this comparison, we approximate how the rate of BBH mergers would be affected and identify qualitative differences in the resulting populations.
In Sec.~\ref{sec:methods} we describe our comparison and the stellar and binary assumptions made. 
In Sec.~\ref{sec:simulations} we describe our model variations, and in Sec.~\ref{sec:MESA_results} we show the outcomes of our binary simulations ran with \MESA{}.
In Sec.~\ref{sec:cosmic_and_mesa_gris_comparison} we provide a qualitative comparison between the outcomes found with \MESA{} to the same binary systems simulated with \COSMIC{}. 
In Sec.~\ref{sec:relative_rate_calculation} we present our quantitative comparisons between binary populations ran with \COSMIC{} to those informed by our detailed simulations. 
We find that the dominant formation channel is stable MT across the range of masses we consider, and that the merger times for BBH differ between \COSMIC{} and our detailed simulations with \MESA{}. 
Consequently, merger rates from isolated binary evolution calculated using current population synthesis prescriptions may be significantly overestimated.

\section{Method}\label{sec:methods}

We compare BBH mergers between the rapid population synthesis code \COSMIC{}, and detailed \MESA{} binary evolutionary models. 
\COSMIC{} is based on {{\texttt{BSE}}\xspace{}}, which uses stellar evolutionary models from \cite{Pols1998,Hurley2000,Hurley2002}, and includes some updates to massive star evolution \citep{Breivik2020,Zevin2020}.
\MESA{} is a one-dimensional stellar evolution code that also includes physical prescriptions for binary stellar evolution \citep{Paxton2011,Paxton2013,Paxton2015,Paxton2019}.
In Sec.~\ref{sec:relativeratemethods} we describe how we combine an initial population of binaries from the code \COSMIC{} with detailed grids of \MESA{} simulations to perform this comparison. 
In Sec.~\ref{subsection:star physics} and Sec.~\ref{subsection:binary physics} we describe our stellar and binary physics models and assumptions. 
Since our objective is to assess how detailed modeling of the BH--H-rich star stage affects the final evolutionary outcome, we are careful to maintain consistency in all other areas of stellar and binary physics between the two codes. 
Our simulations are computed using version 12115 of \MESA{}, and version 3.3 of \COSMIC{}.

\subsection{Relative Rate Calculation} \label{sec:relativeratemethods}

We use \COSMIC{} to generate an initial population of binaries formed from a single burst of star formation with binary parameters initialized following \citet{MoeDiStefano2017}. 
We evolve these binaries from zero-age MS (ZAMS) until the formation of a hydrogen-rich donor with a BH companion (BH--H-rich star). 
To compare binary evolution for different donor masses, we create subpopulations by selecting systems where the H-rich donor is in a specified mass range, e.g., $M_{\rm donor}= (25 \pm 2.5) M_{\odot}$, and with a mass ratio $q = M_{\rm accretor} / M_{\rm donor}$, $q < 1$.  
We consider four subpopulations of BH--H-rich star binaries with different donor mass ranges: $M_{\rm donor}= (25 \pm 2.5) M_{\odot}$, $(30 \pm 2.5) M_{\odot}$, $(35 \pm 2.5) M_{\odot}$, and $(40 \pm 2.5) M_{\odot}$. 

We compute the subsequent evolution of the BH--H-rich star binaries with both \COSMIC{} and \MESA{}. 
Using \COSMIC{}, we continue the evolution of the binaries from our initial population. 
For \MESA{}, instead of simulating each unique binary system within our \COSMIC{} subpopulation, we generate a grid of BH--MS binaries with varying mass ratio and initial orbital period for each of our four subpopulations. 
This means that we select a \emph{single} donor mass in \MESA{} to compare to a selected mass range of \COSMIC{} systems (i.e., a mass range of $M_{\rm donor} = (30 \pm 2.5)$ $M_{\odot}$ in our \COSMIC{} models is compared to a single grid of \MESA{} simulations with $M_{\rm donor} = 30 M_{\odot}$). We also simplify all H-rich stars in \COSMIC{} as MS stars in \MESA{}.

Given the final evolutionary state of the binaries modeled with \COSMIC{} and \MESA{}, we can identify differences in the orbital period and mass ratio for BH--H-rich star binaries that evolve to form merging BBHs.
Combining this information with the number of BH--H-rich star binaries formed in the initial \COSMIC{} population, we can estimate the number of merging BBHs. 
Using these results, we calculate a \emph{\numberratio{}} $\mathcal{N}$ for each donor mass,
\begin{equation}
    \mathcal{N}(\tau) = \frac{\nmergercosmic(\tau)}{\nmergermesa(\tau)},
    \label{eq:relative-rate}
\end{equation} 
where $\nmergercosmic(\tau)$ is the number of BBHs that merge at time $\tau$ resulting from binary systems evolved with \COSMIC{}, and $\nmergermesa(\tau)$ is the number of BBH mergers following the BH--MS evolution with \MESA{} including its detailed treatment of MT. 
We will primarily consider the number of mergers within a Hubble time ($\tHubble = 13.8~\mathrm{Gyr}$) and the ratio $\mathcal{N}(<\tHubble)$.

While $\nmergercosmic$ can be calculated by counting the number of BBH mergers in the \COSMIC{} simulation, to calculate $\nmergermesa$, we must combine the initial BH--H-rich star subpopulations from \COSMIC{} and our grid of \MESA{} models. 
For each donor mass range, we:
\begin{enumerate}
    \item Bin the subpopulation of \COSMIC{} BH--H-rich star systems in mass ratio and orbital period to compare to our \MESA{} models. 
    The size of bins is set to match the highest resolution of our \MESA{} grids so that each bin corresponds to one \MESA{} simulation.
    \item  Identify all bins where the outcome of their corresponding \MESA{} simulation results in a BBH that merges within a Hubble time. 
    \item Assume that all of the \COSMIC{} BH--H-rich systems evolved within these bins will result in a BBH merger. 
    \item Sum the total number of binaries in those bins. 
\end{enumerate}
The final sum is $\nmergermesa(<\tHubble)$, the number of BBH mergers resulting from the initial \COSMIC{} population, but except now BH--H-rich star MT and CE is treated using \MESA{} (see Sec.~\ref{sec:relative_rate_calculation} for an illustration of this procedure). 
We have verified that the resolution of our \MESA{} grid does not significantly influence our results.

To illustrate the difference in merger times, we calculate an estimated merger rate from a subpopulation
\begin{equation}
   \mathcal{R}_\mathrm{mgr} = \sum_{i\,=\,1}^{N_\mathrm{mrg}(<\tHubble)} \frac{1}{t_{\mathrm{mrg},i}} = N_\mathrm{mrg}(<\tHubble)\left\langle\frac{1}{t_\mathrm{mrg}}\right\rangle,
\end{equation}
where the sum is over all binaries that merge within a Hubble time, the weighting is by the inverse of the binary's merger time $t_{\mathrm{mrg},i}$, and $\langle1/t_{\mathrm{mrg}}\rangle$ is the mean inverse merger time for the population. 
We use this to calculate a \emph{\relativerate{}}, 
\begin{equation}
    \mathR = \frac{\Rcosmic}{\Rmesa}.
\end{equation}
The \relativerate{} \mathR{} does not directly correspond to the expected ratio of astrophysical merger rates. 
Calculating these requires simulating an evolving population with varying star formation and metallicity across the history of the Universe. 
This is beyond the scope of this study. 
However, if \mathR{} is close to $1$, we expect the astrophysical rates predicted using \COSMIC{} and detailed simulations to be similar, and the differences in modeling to have negligible impact, whereas a value of \mathR{} different from $1$ indicates a discrepancy in predictions.

\subsection{Stellar Physics} \label{subsection:star physics}

The stellar physics in our \MESA{} simulations is similar to what is implemented in the \texttt{MESA Isochrones and Stellar Tracks} library \citep[\texttt{MIST};][]{Choi2016}. 
Our standard models are initialized at a metallicity of $Z = 0.1 Z_{\odot}$, defining $Z_{\odot} = 0.0142$ and $Y_{\odot}=0.2703$ \citep{2009Asplund}. 
We specify the helium fraction as $Y = Y_\mathrm{Big\ Bang} +  ( Y_{\odot} - Y_\mathrm{Big\ Bang} ) Z/Z_{\odot}$, where $Y_\mathrm{Big\ Bang} = 0.249$ \citep{Planck2016}. 
Nuclear reaction rates are drawn from the JINA Reaclib database \citep{Cyburt2010}. 
We use the \texttt{basic.net} nuclear reactions networks for H and He burning; \texttt{co$\textunderscore$burn.net} for C and O burning, and \texttt{approx21.net} for later phases.  
We use the standard equation of states in \MESA{} of OPAL \citep{Rogers2002}, HELM \citep{Timmes2000}, PC \citep{Potekhin2010}, and SCVH \citep{Saumon1995}. 
Radiative opacities are taken from \citet{Iglesias1996}, which include the impact of enhanced CO-mixtures in the opacities, as well as the opacity calculations of \citet{Ferguson2005} for low temperatures.
Models are computed until they reach core carbon depletion (central $^{12}$C abundance $< 10^{-2}$); at this point we assume that stars will undergo direct core collapse to a BH with mass equal to their baryonic mass.

Stars are assumed to be synchronized with the orbital period at the beginning of the \MESA{} simulations. 
The implementation of rotation in \MESA{} closely follows \citet{Heger2000,Heger2005}. 
We include rotational mixing with an efficiency parameter of $f_{\rm c} = 1/30$ \citep{Heger2000,Chaboyer1992}. 
We include transport of angular momentum by magnetic fields due to the Spruit--Tayler dynamo \citep{Spruit2002}. 

For convective mixing, we adopt the mixing-length theory of \citet{Mihalas1978book} and \citet{Kurucz1970} with a convection mixing-length parameter $\alpha_{\rm MLT} = 1.93$ based on results
from the \texttt{MIST} project (Dotter et al., in preparation). 
Convective boundaries are determined with the Ledoux criterion. 
We do not include semi-convection mixing and instead use convective premixing \citep[][section 5.2]{Paxton2019}.  
\MESA{} treats thermohaline mixing with a diffusion coefficient motivated by \citet{Ulrich1972} and \citet{Kippenhahn1980}.  
We adopt a thermohaline mixing efficiency parameter of $\alpha_{\rm th} = 17.5$, which follows \citet{Charbonnel2007} with an aspect ratio of $1$ for the instability fingers.  
Overshoot mixing is treated with the exponential decay formalism \citep{Herwig2000,Paxton2011}. 
For the range of stellar masses studied here, we use $f_{\rm ov}=0.0415$, which describes the extent of the overshoot mixing in this formalism.
This value is motivated by \citet{Brott2011}, who used the step overshoot formalism, and the work of \citet{Claret2017} is used to translate the step to exponential decay overshoot. 
To prevent numerical issues caused by radiation-dominated envelopes, such as those in massive stars, we use the \texttt{MLT++} treatment of convection \citep[][section 7.3]{Paxton2013}. {\highlight While this treatment helps prevent numerical issues, it reduces the radius expansion and can affect MT for massive donors \citep{Klencki2020a}. 
The use of the \texttt{MLT++} treatment is in contrast to \cite{Marchant2021}, but (as discussed in Sec.~\ref{sec:MESA_results}) we find consistent results, indicating that our qualitative conclusions are not significantly impacted by this choice.}

For stellar winds we use the \texttt{Dutch} prescription in \MESA{}, which is based on \citet{Glebbeek2009}. 
It uses \citet{Vink2001_hot_highH} for effective temperatures of $T_{\rm eff} > 10^4~\mathrm{K}$ and surface H mass fraction of $H>0.4$; \citet{Nugis2000_hot_lowH} for $T_{\rm eff} > 10^4~\mathrm{K}$ and $H<0.4$ (Wolf--Rayet stars), and \citet{deJager1988} for $T_{\rm eff} < 10^4~\mathrm{K}$. 

The \COSMIC{} wind prescription most similar to the \texttt{Dutch} prescription treats O and B stars following \citet{Vink2001_hot_highH}, and Wolf--Rayet stars following \citet{Hamann1998} reduced by factor of $10$ \citep{Yoon2010} with metallicity scaling of $(Z/Z_{\odot})^
{0.86}$ \citep{Vink2005}. 
We expect the differences between the winds used for \MESA{} and \COSMIC{}  to not significantly affect our results.

In \COSMIC{}, instead of direct core collapse, we follow the delayed prescription of \citet{Fryer2012}. 
{\highlight Other models like the probabilistic prescription of \citet{Mandel2020} or based upon the simulations of \citet{PattonSukhbold2020} indicate that the delayed prescription of \cite{Fryer2012} may overestimate the amount of mass lost during the formation of the BH, resulting in lower-mass BHs \citep[e.g.,][]{Patton2021}. 
While this may have an impact on the GW inspiral times, it should not significantly affect the relative contribution of stable and unstable MT for BBH mergers. }
For both models ran with \MESA{} and \COSMIC{} we do not introduce supernova (SN) kicks. 
However, our \COSMIC{} models still have kicks from mass loss. The differences in SN prescriptions do not weaken our final results since the direct core-collapse method is on the more optimistic side of producing BBH mergers.

\subsection{Binary Evolution Physics} \label{subsection:binary physics}

In order to enable a more direct comparison between \MESA{} and \COSMIC{} we make an effort to maintain consistency in all other parts of binary evolution modeling. 
However, notable differences in the computation of binary evolution between the two codes still exist. 
The primary differences stem from the ability of \MESA{} to model binary evolution in more detail.

For both \COSMIC{} and \MESA{}, we initialize BH--MS binary systems with non-spinning BHs but, unlike \COSMIC{}, \MESA{} evolves the spin-up of the BH through accretion \citep{Marchant2017}. 

To compute the effect of tides in \MESA{} we apply a structure-dependent tidal torque summed throughout the whole star. Equilibrium tides are applied relative to convective zones \citep{Hut1981,Hurley2002} and dynamical tides connected to radiative zones. We use the tidal coefficient $E_2$ calculated in \citet{Qin2018}, based upon \citet{Zahn1977}, to calculate the dynamical tides.
Tides modeled in \COSMIC{} follow \citet{belczynski_compact_2008}, which includes either equilibrium tides or dynamical tides based strictly upon the stellar type and mass \citep{Hurley2002}.

In both \COSMIC{} and \MESA{} we assume conservative MT for sub-Eddington accretion onto a BH. 
In our standard model, we limit accretion to the Eddington limit and explore accretion limits up to $10$ times the Eddington limit. 
We allow for Bondi--Hoyle accretion \citep{BondiHoyle1944} and adopt an efficiency factor $\alpha_{\rm BH} = 1.5$ as in \citet{Hurley2002}.  
Mass lost during super-Eddington MT is removed from the system with the specific angular momentum of the accretor. 

BH--MS models ran with \MESA{} remain in circular orbits during binary evolution while binary systems in \COSMIC{} may be born with or gain eccentricity during kicks. 
However, our comparison begins at the BH--H-rich stage when the eccentricities of \COSMIC{} systems are relatively low ($e \lesssim 0.2$). 
We have confirmed that the additional orbital angular momentum of these eccentric binaries, when translated to a corresponding circular orbit, did not significantly affect the results. 
For both codes the evolution of orbital angular momentum considers the effects of mass loss, gravitational radiation \citep{Peters1963}, and tides as described above.

\subsubsection{Common-envelope Evolution } \label{subsection:CE}
A crucial difference between binary evolution modeled in \COSMIC{} and \MESA{} is how CE is initiated and treated. 
In \COSMIC{}, CE evolution follows the classic $\alpha_{\rm CE}$--$\lambda$ prescription \citep{Webbink1984,deKool1990,DewiTauris2000} with critical mass ratio $q_{\rm crit}$ values to determine the onset of CE. In \MESA{}, we implement the detailed model for MT and CE evolution following the method of \cite{Marchant2021}.

In our \COSMIC{} simulations we use the variable $\lambda$ prescriptions by \citet{Claeys2014} set to include only gravitational potential energy to unbind the envelope $\lambda_\mathrm{g}$ \citep{DewiTauris2000}. 
Since the default behavior of the \cite{Claeys2014} $\lambda$ prescription includes both gravitation and thermal energy, where it is approximated as $\lambda = 2\lambda_\mathrm{g}$, we scale the $\lambda$ value by a factor of $1/2$. 
Our standard \COSMIC{} models are ran with $q_{\rm crit}$ following \citet{belczynski_compact_2008}. 
We also assume the optimistic CE scenario where systems with a Hertzsprung gap star can survive the CE phase \citep{Belczynski2010}. 
As a comparison we also use the pessimistic scenario, where these same systems do not survive.

In the method of \citet{Marchant2021}, 
the prescription for MT is an extension of the one developed by \citet{Ritter1988} and \citet{KolbRitter1990}, and accounts for cases of large overflow as well as potential outflows from outer Lagrangian points. 
To model the outcome of CE evolution we follow a similar method to the usual $\alpha_\mathrm{CE}$--$\lambda$ prescription, but aim to determine the binding energy self-consistently with the stellar model. 
Whenever a donor exceeds a MT rate of $\dot{M}> 1 M_{\odot}~\mathrm{yr}^{-1}$ we consider MT to be unstable, and model the evolution as a CE system. 
At the onset of instability, we compute the binding energy outside an arbitrary mass coordinate $m$,
\begin{equation}
    E_\mathrm{bind} (m) =  \int^{M_{\rm donor}}_{m} \left( - \frac{G m'}{r} + \alpha_{\rm th} u\right) {\rm d}m',
\end{equation}
where $u$ is the specific internal energy of the gas and $\alpha_\mathrm{th}$ represents the fraction of this energy that can be used for the ejection of the envelope. 
For consistency with the assumptions of our \COSMIC{} models, we take $\alpha_\mathrm{th}=0$.
Given $E_\mathrm{bind}$ and the efficiency $\alpha_\mathrm{CE}$, one can determine the post-CE orbital separation if one knows the mass coordinate $M_\mathrm{core}$ at which the CE process completes; to determine this, we simulate a rapid mass-loss phase from the star while updating the separation of the binary using the binding energy at each mass coordinate, until the system detaches \citep[][section 2.2]{Marchant2021}. 

\section{Model Variations} \label{sec:simulations}

The configurations in our standard \COSMIC{} and \MESA{} models are for metallicity $Z =0.1 Z_{\odot}$ and an effective CE efficiency of $\alpha_{\rm CE} = 1$. 
Parameters specific to \COSMIC{} in our standard model include $\lambda$ prescriptions by \citet{Claeys2014} and $q_{\rm crit}$ following \citet{belczynski_compact_2008}.

In addition to the standard model, we consider five additional models:
\begin{enumerate}
    \item {$Z$} -- We simulate systems at solar metallicity.
    
    \item \emph{accretion} -- Simulations have shown that mass accretion onto a BH can exceed the Eddington limit \citep{Begelman2002, McKinney2014}.
    {\highlight Super-Eddington accretion has also been studied in the context of massive BH binaries in the pair-instability mass gap \citep{vanSon2020}.}
     We thus allow for super-Eddington accretion at $10$ times the Eddington limit. 
    
    \item $\alpha_{\rm CE}$ --  We apply a higher CE efficiency of $\alpha_{\rm CE} = 5$. 
    Three-dimensional hydrodynamical studies of successful envelope ejections have suggested $\alpha_{\rm CE}$ values as low as $0.1$--$0.5$ for extended red supergiant progenitors \citep{LawSmith2020}. 
    However, some one-dimensional simulations of CE evolution that include additional energy sources like ionization and thermal energy suggest a high CE efficiency of $5$ \citep{Fragos2019}. 
    In addition to this many studies have previously concluded that $\alpha_{\rm CE}$ values up to $5$ are at least marginally preferred in order to match the GW observations  \citep{Giacobbo2018,Santoliquido2021,Zevin2021}. 
    We therefore consider this higher CE efficiency in order to explore the more optimistic CE assumption. 
 
    \item $q_{\rm crit}$ -- Based on the system's mass ratio at the onset of Roche lobe overflow (RLOF), the stability criteria, $q_{\rm crit}$, determines if a system will undergo a CE event or if it will proceed with stable MT. 
    {\highlight In our variation we use the $q_{\rm crit}$ prescription from \cite{Claeys2014}, which allows more systems with a Hertzsprung gap donor to proceed with stable MT instead of CE.
    Instead of $q_{\rm crit}=0.33$ for all H-rich donors as in our standard configuration,
    \cite{Claeys2014} uses $q_{\rm crit}=1$ for MS stars, $q_{\rm crit}=0.21$ for Hertzsprung gap stars, $q_{\rm crit}=0.33$ for core helium burning donors, and $q_{\rm crit}=0.87$ for first giant branch stars, early asymptotic giant branch (AGB), and thermally pulsing AGB. Changing the $q_{\rm crit}$ prescription also introduces a change to the MT rates \citep{Breivik2020}. This variation also uses MT rates consistent with \citet{Claeys2014}, which implements higher MT rates during thermal-timescale MT.}

    \item \emph{CE} -- We apply the Pessimistic scenario that causes a subset of stellar types to merge during the CE phase \citep{Belczynski2007_pess}. 
    In \COSMIC{} these stellar types include MS, {\highlight Hertzsprung gap,} naked helium {\highlight MS, naked helium Hertzsprung gap}, and white dwarfs.
    However, recent work shows that the survivability of CE is more attributed to the type of envelope of the donor rather than the evolutionary type of the donor \citep{Klencki2021,Marchant2021}, and the type of envelope cannot be linked to specific stellar types \citep{Klencki2020a}. Therefore, the Pessimistic model for CE used here is likely not restrictive enough. 
\end{enumerate}
The first three variations require generating new grids of \MESA{} models along with new \COSMIC{} populations, while the final two are specific to \COSMIC{} (and other population synthesis codes). 
We compare these variations in \COSMIC{} to our standard \MESA{} model.  

\section{MESA Results} \label{sec:MESA_results}

\begin{figure*}[!htb]
\centering
\includegraphics[width=0.7\textwidth]{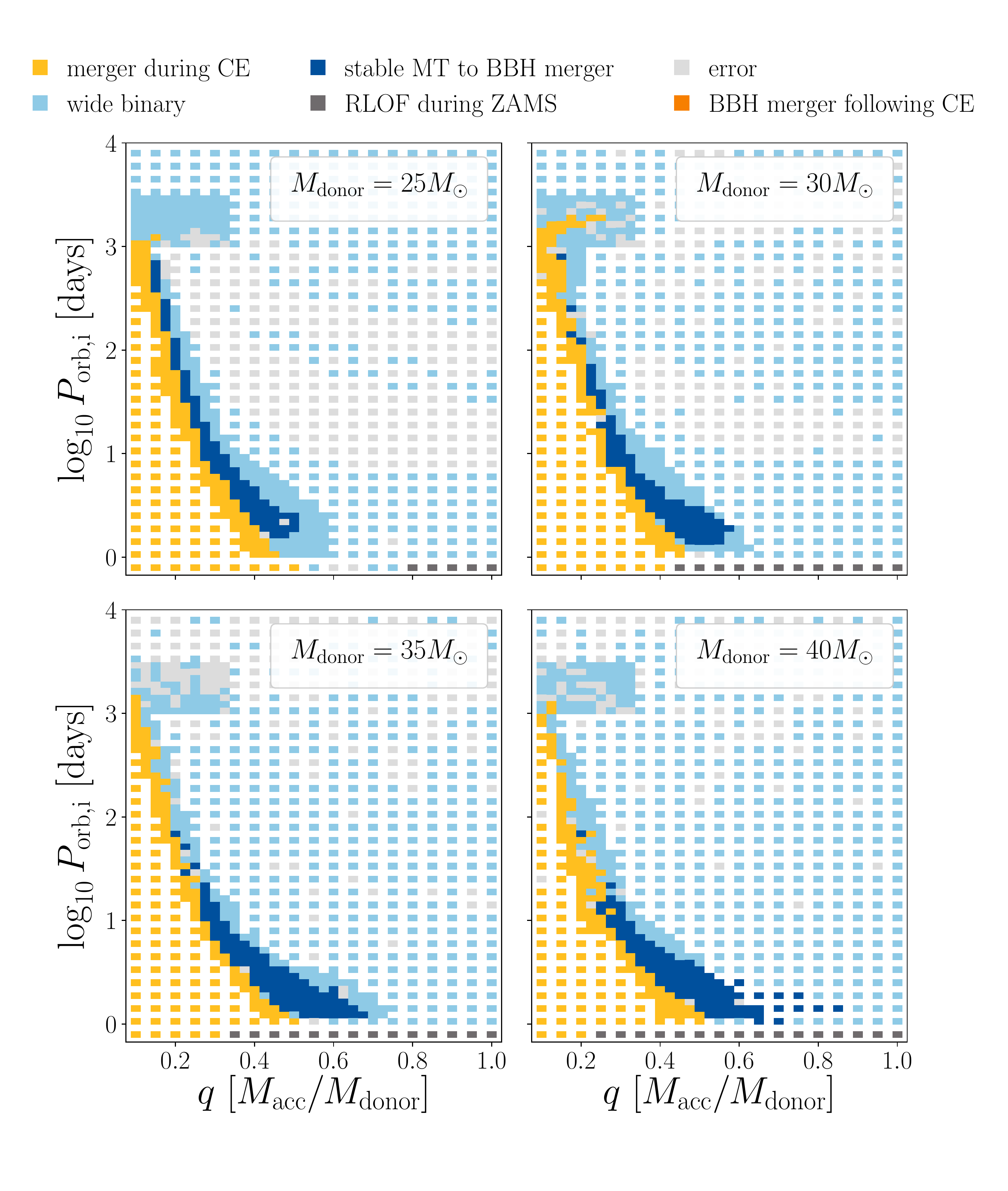}
\caption{ Final outcomes of our standard models of BH--MS systems ran with \MESA{}. 
They have sub-solar metallicity $Z = 0.1 Z_\odot$, CE efficiency $\alpha_{\rm CE} = 1$ and no additional sources of energy into unbinding the envelope, $\alpha_{\rm th} = 0$. 
Each panel corresponds to a different donor mass. 
For all masses, we find a narrow band of systems whose outcome results in a BBH merger within a Hubble time following only a phase of stable MT (dark blue). 
For these models we do not find any successful CE ejections.} 
\label{fig:standard_grid_results}
\end{figure*}

\subsection{Standard Model}
{ \highlight We computed four grids of models, one for each donor mass $25 M_{\odot}$, $30 M_{\odot}$, $35 M_{\odot}$, and $40 M_{\odot}$, consisting of a MS donor with a BH accretor in a circular orbit.
The grids span a period range between $-0.1  < \log_{10}(P_{\rm orb,i}/{\rm days})< 4$, and initial mass ratios $q=M_{\rm accretor}/M_{\rm donor}$ between $0.1$ and $1$. }
Figure~\ref{fig:standard_grid_results} shows the final outcomes of our standard model of \MESA{} simulations as a function of initial mass ratio $q$ and initial orbital period. 
Each panel in Fig.~\ref{fig:standard_grid_results} corresponds to a different donor mass. 
At these masses, the outcomes show a consistent picture. 
The (yellow) region corresponding to merger during CE occupies the bottom left of these plots. 
The region where systems undergo CE evolution is roughly the same for all donor masses: it extends from low-period orbits with $q\simeq 0.5$ diagonally upward to $P_{\rm orb, i} \simeq  1000~\mathrm{days}$ and $q\simeq 0.15$. 
In addition to this region, we also find a hook of CE evolution for $M_{\rm donor} = 30 M_{\odot}$ at $P_{\rm orb, i} \simeq 1000~\mathrm{days}$. 
For all donor masses with our standard model we do not find any successful CE ejections, and hence there are no BBHs formed from CE evolution. 
However, we do find a (dark blue) region where the final outcome is a BBH merger within a Hubble time following a phase of stable MT. 
This region forms a narrow transition between systems merging during CE and systems that end up as wide, {\highlight non-merging BBHs (light blue).  We do not distinguish wide binaries with stable MT from wide binaries with no interaction.}
As donor mass increases, the region corresponding to BBH merger following only stable MT shifts to lower initial orbital periods and extends to larger initial mass ratios. 
In all cases we find that the only formation channel for BBH mergers is stable MT. In Sec.~\ref{sec:cosmic_and_mesa_gris_comparison} we compare the grids ran with $M_{\rm donor} = 25M_{\odot}$ and $40M_{\odot}$ to grids produced with \COSMIC{}. 

The grid of systems with $M_{\rm donor} = 30 M_{\odot}$ can be compared to the results in \citet{Marchant2021}. 
Both show a hook of CE evolution at $P_{\rm orb,i}\approx1000~\mathrm{days}$. 
This hook of CE evolution in our grid is smaller, and unlike \citet{Marchant2021} this region of CE evolution does not result in any successful ejections. 
A key difference that affects the survivability of CE between \cite{Marchant2021} and this work is our exclusion of the thermal energy and recombination component in CE ejection. 
However, when including thermal and recombination energy \citet{Marchant2021} did not find a significant number of successful CE ejections leading to merging BBH. They find that under the assumption of a flat distribution in mass
ratio and a flat distribution in $\log_{10} P_{\rm orb, i}$ for their grid with $M_{\rm donor} = 30M_{\odot}$, the ratio of the number of CE simulations that produce merging BBHs to the number of stable MT simulations that produce merging BBHs is $0.017$. 
Similar outcomes have also been shown under more extreme scenarios. \citet{Klencki2021} found that even under a set of optimistic assumptions about CE evolution, a successful CE ejection is only possible when the donor has a massive convective envelope. 

Another feature in our results that can be directly compared to \cite{Marchant2021} is the (dark blue) region corresponding to BBH mergers following stable MT. 
Both this work and \cite{Marchant2021} show that the dominant formation channel for BBH mergers is through stable MT, and that this region becomes more narrow with increasing initial orbital period. 
Additionally, \citet{Marchant2021} found that the boundary between stable and unstable MT was robust under different thresholds of unstable MT. 
These results are in agreement with previous studies. 
\cite{ge_adiabatic_2015} also found a similar trend for the boundary between stable and unstable MT: the boundary allows for more stability with increasing stellar age (increasing initial orbital period in our grids). 
Using stability criteria based on the work of \citet{ge_adiabatic_2015} for MS and Hertzsprung gap stars, \citet{neijssel_effect_2019} found that $80\%$ of the BBH mergers are formed from stable MT, not CE.
Similarly, \citet{olejak2021impact} used criteria for the onset of CE derived from the simulations of \citet{Pavlovskii2017} to find that BBH formation could be dominated by stable MT without a CE, although this conclusion also depended upon the MT timescale. 
\citet{Shao2021} combined their \texttt{BSE} population synthesis with MT stability criteria derived from grids of simulations ran with \MESA{}. 
They assume that a CE is initiated if the MT rate exceeds a critical value based upon either (i) the Eddington rate at the photon-trapping radius \citep{Begelman1979,King1999,belczynski_compact_2008}, or (ii) $2\%$ of the donor mass per orbit of the donor overflows the second Lagrange point \citep{Pavlovskii2015,Ge2020}, and find that $\sim30$--$70\%$ of merging BBHs form from stable MT.
These results highlight that CE evolution is not needed to form merging BBHs.

These models at $Z=0.1 Z_{\odot}$ with $M_{\rm donor} = 30 M_{\odot}$,  $35 M_{\odot}$, and $40 M_{\odot}$ experience numerical issues after the depletion of core hydrogen. 
The numerical issues cause the star's radius to vary and core mass to increase but the donor regains stability and continues its post-MS evolution. 
The discontinuities seen in the (dark blue) region corresponding to BBH mergers following stable MT may be attributed to these numerical issues. 
However, we do not expect such issues to have significant affects on our main conclusions or the qualitative behavior across parameter space. 
{\highlight Additionally, about $30\%$ of the errors in the grid with $M_{\rm donor} = 25 M_{\odot}$ occurred close to core carbon depletion. Since these occur outside our regions of interest, and do not influence our results, we did not resolve the errors to continue the evolution to carbon depletion. }

\subsection{Variations to Standard Model} \label{subsection:mesa grid results, variations to standard}

\begin{figure*}[hbt!]
\centering
\includegraphics[width=0.7\textwidth]{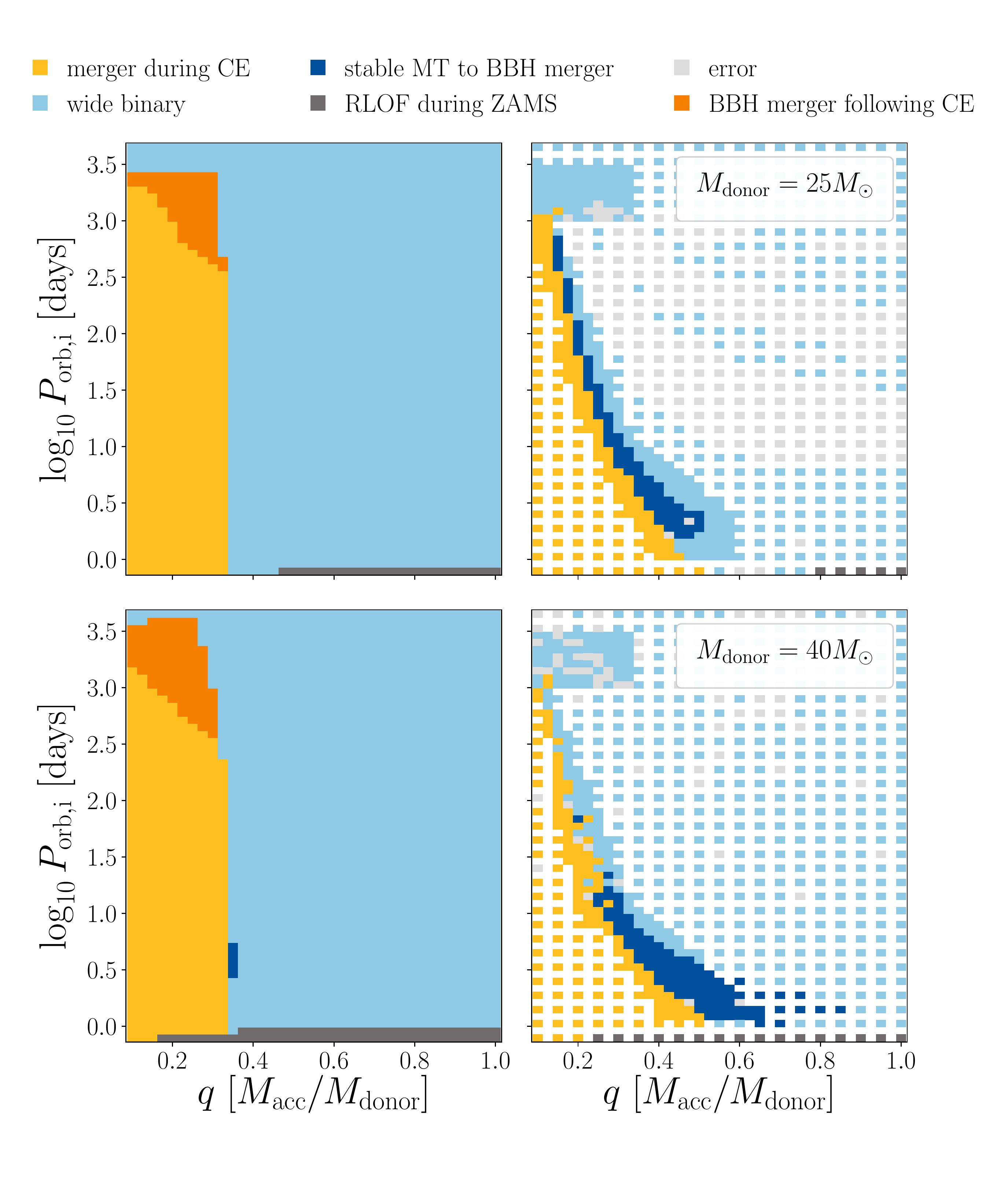}
\caption{Comparison of final outcomes between binary simulations ran with \COSMIC{} ({\it left}) to simulations ran with \MESA{} ({\it right}). {\it Top}: comparison for MS--BH binaries with $M_{\rm donor} = 25 M_{\odot}$ shown as a function of initial orbital period and mass rations, $q$. 
{\it Bottom}: same as top but for $M_{\rm donor} = 40 M_{\odot}$. 
For binary evolution modeled with \COSMIC, CE is the dominant formation channel for BBH mergers. 
For evolution modeled with \MESA{} we do not find any successful CE ejections and all BBH mergers are formed through stable MT. } 
\label{fig:COSMIC_MESA_grid_comparison}
\end{figure*}

As described in Sec~\ref{sec:simulations}, we generated three more sets of grids where we varied the CE efficiency $\alpha_{\rm CE}$, allowed for super-Eddington MT, and changed to solar metallicity.

Increasing the CE efficiency from $\alpha_{\rm CE}=1$ to $\alpha_{\rm CE}=5$  (maintaining $\alpha_{\rm th}=0$) results in little difference overall. 
However, for the grid with $M_{\rm donor} = 25 M_{\odot}$, we find three simulations that result in a BBH merger following a successful CE ejection. 
These binaries are on the boundary between systems merging during the CE phase (yellow region) and simulations resulting in a BBH merger following stable MT (dark blue region). 
For $M_{\rm donor} = 30 M_{\odot}$, we find successful CE ejections in the hook of CE evolution at $P_{\rm orb, i} \approx 1000~\mathrm{days}$. 
For $M_{\rm donor} = 35 M_{\odot}$, we do not find a difference when using $\alpha_{\rm CE}=5$. 
For the grid of simulations with $M_{\rm donor} = 40 M_{\odot}$ we find one system resulting in a BBH merger following CE. 

Allowing for accretion at $10$ times the Eddington limit, the regions where systems undergo a CE phase are similar to our standard models. 
We find more BBH mergers following stable MT at low initial orbital periods, roughly centered at low initial orbital periods $P_{\rm orb, i}\simeq 100~\mathrm{days}$. 

We find the greatest differences for the \MESA{} models initialized at solar metallicity $Z = 0.0142$. 
Overall, we find fewer BBH mergers following stable MT and find fewer binaries undergoing CE. 
This is likely for two main reasons. 
First, strong stellar winds for solar metallicity widen the orbits compared to $Z=0.1 Z_{\odot}$. 
Therefore, fewer systems merge within a Hubble time and we only find CE evolution for tighter initial orbital periods. 
Second, the radii of these stars are larger, causing RLOF during ZAMS at higher orbital periods. 
The only exception to an overall decrease in CE systems is for $M_{\rm donor} = 25$ and $40 M_{\odot}$. 
In both of these cases, we find a hook of CE evolution at $P\sim 1000~\mathrm{days}$, similar to the feature in Fig.~\ref{fig:standard_grid_results}.

\section{COSMIC and MESA comparison} \label{sec:cosmic_and_mesa_gris_comparison}

Before combining \MESA{} and \COSMIC{} together to calculate the relative rates, we first compare BH--MS evolution from \MESA{} to that with \COSMIC{}. 
Although our relative rate calculation involves all BH--H-rich star binaries in \COSMIC{}, including BH--MS, 
the differences presented here propagate into the relative rate calculation, as they determine the regions of parameter space where merging BBHs can form. 

Figure~\ref{fig:COSMIC_MESA_grid_comparison} shows the final outcomes of simulations for BH--MS binaries using our standard model. 
There are several key differences between our \MESA{} and \COSMIC{} simulations. 
First, we focus on outcomes involving CE evolution. 
We find that models ran with \COSMIC{} develop strong interactions (e.g., CE interactions) at higher orbital periods ($P_{\rm orb, i} \sim 3000$~days) than models ran with \MESA{} ($P_{\rm orb, i} \sim 1000$~days).
The most frequent outcome are systems that result in a merger during the CE phase. 
In \COSMIC{}, this (yellow) region extends from high initial orbital periods $P_{\rm orb,i} \sim 1000~\mathrm{days}$ at our lowest $q$ to $P_{\rm orb,i} \sim 300~\mathrm{days}$ where it sharply cuts off at $q \sim 0.3$. 
This sharp boundary is a result of the $q_{\rm crit}$ prescription used in \COSMIC{} to determine stable and unstable MT \citep{belczynski_compact_2008}. 
For models ran with \MESA{}, the same boundary between systems merging during CE and stable MT expands up to similar orbital periods as \COSMIC{}, but instead of a sharp cutoff, the boundary sweeps from low $q$ at high orbital periods gradually to $q \simeq 0.5$ at low orbital periods. MT stability increases with initial orbital period. 
In \COSMIC{} we also find systems resulting in a BBH merger following a successful CE ejection. 
This (orange) region is isolated at $q \lesssim 0.3$ between $P_{\rm orb, i} \approx 300$--$3000~\mathrm{days}$. 
With our standard \MESA{} model, we do not find any simulations that result in a BBH merger following CE. 
Moreover, the region where \COSMIC{} binaries result in BBH mergers following CE is predicted to be stable MT for binaries modeled with \MESA{}. 
Thus, not only are BBH mergers following a successful CE ejection missing in \MESA{}, most of this region does not develop unstable MT. 

A second channel for BBH mergers is through stable MT (dark blue region). 
For the standard \COSMIC{} models, this formation channel only occurs in the bottom panel with $M_{\rm donor} = 40 M_{\odot}$. 
It is a small set of systems at $q \sim 0.35$ and $P_{\rm orb, i} \sim 5~\mathrm{days}$. 
In contrast, this outcome occurs for a significant number of systems in the \MESA{} grids. 
Most of these mergers occur at the boundary between stable and unstable MT, in a band that spans between $q \sim 0.2$--$0.9$, and widens with smaller initial orbital period. 

Differences in the systems undergoing MT are expected because of (i) differences in stellar radius between the different stellar models \citep{Agrawal2020}, and (ii) differences in how MT impacts stellar structure. 
The difference in radii influences which stars undergo MT when. 
The differences in stellar structure arise because \COSMIC{} effectively models the donor as a single star, with mass loss changing the mass and age of this single star \citep[][section 7.1]{Hurley2000}. However, with detailed modeling, the structure of a star that has undergone MT is distinct from a single star of a comparable mass \citep[e.g.,][]{Laplace2021}. 
The addition of BBHs forming through stable MT mitigates some of the loss from CE evolution when comparing the total number of merging BBHs from our \COSMIC{} population and the population simulated using \MESA{}.

\section{Relative Rate Calculation} \label{sec:relative_rate_calculation}

\begin{figure*}[hbt!]
\centering
\includegraphics[width=0.75\textwidth]{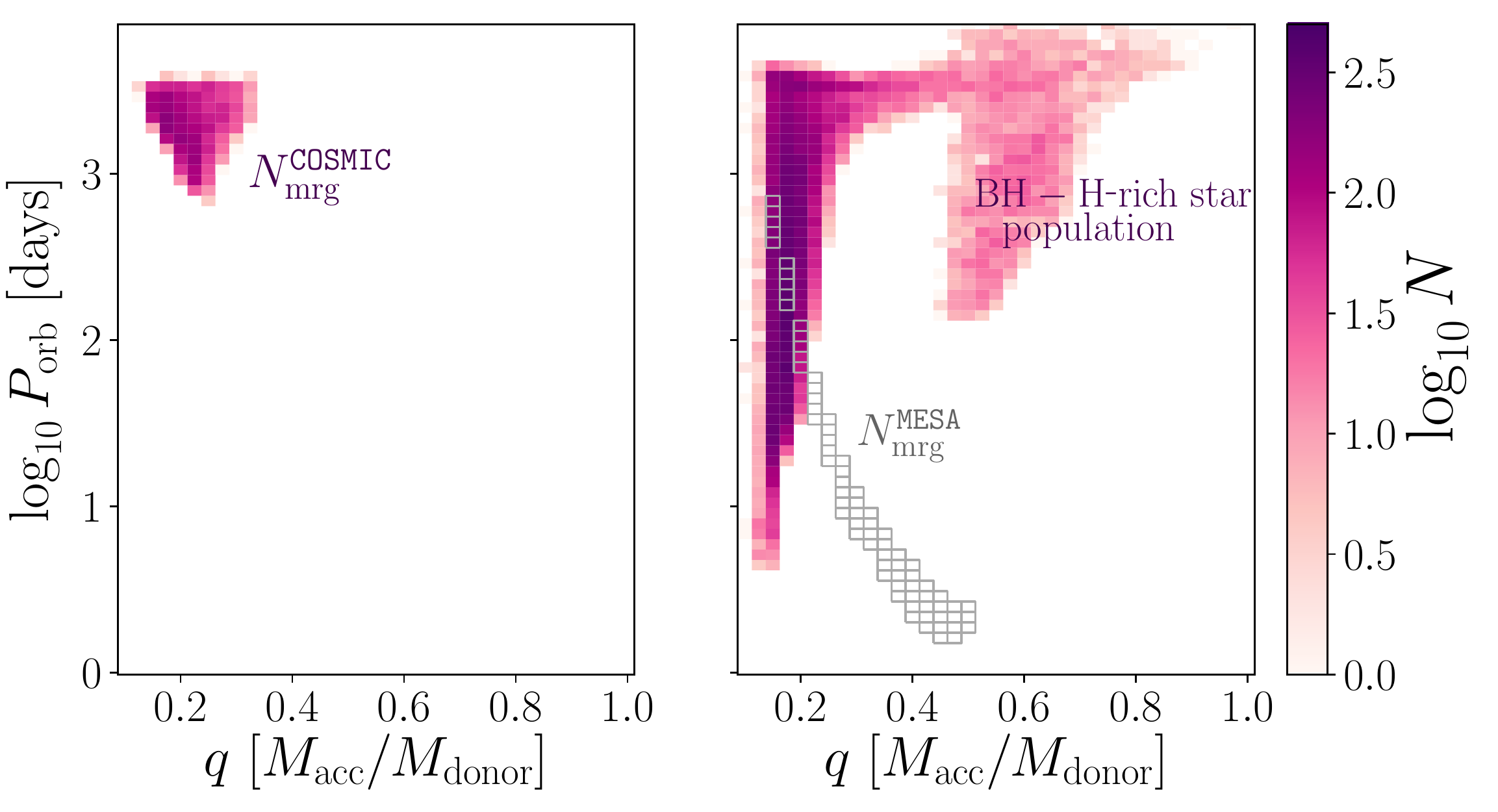}
\caption{Method illustration for donor mass $M_{\rm donor} = (25 \pm 2.5) M_{\odot}$. 
\emph{Left:} Two-dimensional histogram of the final BBH merger population predicted by \COSMIC{} as a function of mass ratio and orbital period when the system became a BH--H-rich star. 
The total number of binaries in this histogram is \nmergercosmic$(<\tHubble)$. 
\emph{Right:} Two-dimensional histogram of BH--H-rich star population. 
Bins outlined in gray are regions where models ran with \MESA{} result in BBH mergers. 
The sum of the number of BH--H-rich star binaries in these gray bins is \nmergermesa$(<\tHubble)$. }
\label{fig: calculation illustration}
\end{figure*}

Figure~\ref{fig: calculation illustration} illustrates how \nmergercosmic{}$(<\tHubble)$ and \nmergermesa{}$(<\tHubble)$ are calculated. 
The left panel in Fig.~\ref{fig: calculation illustration} shows a two-dimensional histogram of binaries from the BH--H-rich star subpopulation with $M_{\rm donor}= (25 \pm 2.5) M_{\odot}$ that results in BBH mergers within a Hubble time in \COSMIC{}. 
We bin the population as a function of $q$ and $P_{\rm orb}$ at the time the BH--H-rich binary was formed. 
The total number of binaries in the left panel is $\nmergercosmic{}(<\tHubble)$.
The right panel shows a two-dimensional histogram of the same BH--H-rich star subpopulation in \COSMIC{} with $M_{\rm donor}= (25 \pm 2.5) M_{\odot}$. 
On this same histogram we outline the bins in gray where BH--MS models ran with \MESA{} result in a BBH merger within a Hubble time. 
The total number of BH--H-rich stars in \COSMIC{} within these bins is \nmergermesa{}$(<\tHubble)$.
The empty outlined bins do not contribute to the total. 
Table~\ref{table:number_ratio_result} shows our results for the \numberratio{} $\mathcal{N}(<\tHubble)  = \nmergercosmic{}(<\tHubble)/\nmergermesa{}(<\tHubble)$ for four different donor masses and six different models.

The first row in Table~\ref{table:number_ratio_result} shows the \numberratio{} for our standard model. 
We find that $\mathcal{N}(<\tHubble)$ is between $\sim1$--$8$ across all donor masses (\COSMIC{} produces up to $8$ times more merging BBHs). 
Another important difference is the dominant formation channel. 
The formation of BBH mergers for our standard \COSMIC{} models form primarily through CE ($\sim99\%$) whereas our \MESA{} models are solely by stable MT at these donor masses (see Fig.~\ref{fig:COSMIC_MESA_grid_comparison}).  

The second row in Table~\ref{table:number_ratio_result} corresponds to the model allowing super-Eddington accretion at $10$ times the Eddington limit. We find that higher accretion rates likely only affect low-period binaries with high accretion rates. 
In addition, Fig.~\ref{fig: calculation illustration} shows that these low-period binaries do not affect our \numberratio{} calculation since they occupy empty period--mass ratio bins.
As a result, the \numberratio{}  with super-Eddington accretion is similar to the standard model. 

The third row in Table~\ref{table:number_ratio_result} corresponds to the model with $\alpha_{\rm CE} = 5$, while maintaining $\alpha_{\rm th} = 0$.
Compared to the standard model and the super-Eddington model, this model with higher CE efficiency results in more varied and higher values of the \numberratio{}. 
For models ran with \COSMIC{}, the value of $\alpha_{\rm CE}$ used here was efficient enough to increase the number of BBH mergers without leading to wider binaries.
On the other hand, for models ran with \MESA{} at this higher $\alpha_{\rm CE}$, the formation of BBH mergers did not change significantly and is still significantly dominated by stable MT.  

The fourth row in Table~\ref{table:number_ratio_result} corresponds to the \numberratio{} for models ran at solar metallicity $Z = Z_{\odot}$. 
For H-rich donors at $M_{\rm donor} = 25 M_{\odot}$ and $30 M_{\odot}$, \COSMIC{} did not produce any merging BBHs. 
For the grids of models with $M_{\rm donor} = 25 M_{\odot}$ and $30 M_{\odot}$ at solar metallicity, all of the BBH mergers found with \MESA{} are outside of the initial \COSMIC{} BH--H-rich star population (unlike the example shown in Fig.~\ref{fig: calculation illustration}). 
Therefore, for these donor masses we cannot calculate $\mathcal{N}$.
At $M_{\rm donor} = 35 M_{\odot}$ and $M_{\rm donor} = 40 M_{\odot}$, we find more merging BBHs with models ran with \MESA{} than \COSMIC{}. 
For $M_{\rm donor} = 35 M_{\odot}$, we can calculate an upper limit (given the finite number of binaries in our population) of $\mathcal{N}(<\tHubble) < 0.06$, and for $M_{\rm donor} = 40 M_{\odot}$ $\mathcal{N}(<\tHubble) = 0.4$.

\begin{deluxetable}{ c c c c c}
\tablecaption{Upper limits of the \numberratio{} $\mathcal{N}(<\tHubble) = \nmergercosmic(<\tHubble)/\nmergermesa(<\tHubble)$ for all model variations and donor masses. The last two rows correspond to variations that are specific to \COSMIC{}. 
We compare these variations in \COSMIC{} to the standard models ran with \MESA{}. From top to bottom the rows correspond to 1) standard model, 2) model allowing for an accretion rate onto the BH at 10 $\times$ the Eddington limit, 3) CE efficiency $\alpha_{\rm CE}=5$ while keeping $\alpha_{\rm th}=0$, 4) solar metallicity, 5) $q_{\rm crit}$ criteria following \cite{Claeys2014}, and 6) using the Pessimistic assumption for CE evolution. 
\label{table:number_ratio_result} }
\tablehead{
\colhead{} & \multicolumn{4}{c}{Subpopulation} \\
\cline{2-5}
\colhead{Model} & $25 M_{\odot}$ & $30 M_{\odot}$ & $35 M_{\odot}$ &  $ 40 M_{\odot}$
}
\startdata 
{Standard} & 1.3 & 4.3 & 7.8 & 2 \\
{$ \dot{M}_{\rm Edd} \times 10$} & 1.7 & 4 & 9 & 1.7 \\
{$\alpha_{\rm CE} = 5$} & 2.6 & 15 & 35  & 8.5 \\ 
{$ Z_{\odot}$ } & - & - & $<$0.06 & 0.4  \\ 
Claeys & 0.7 & 2.7 & 8.5  & 15 \\
Pessimistic & 1.3 & 4.3 & 7.7 & 2 \\
\enddata
\end{deluxetable}

The last two rows in Table~\ref{table:number_ratio_result} correspond to models where we vary parameters specific to \COSMIC{}, and calculate a \numberratio{} with the standard \MESA{} model. 
In the fifth row we change the $q_{\rm crit}$ prescription for MT stability from our standard model using \citet{belczynski_compact_2008} to the prescription of \citet{Claeys2014}. 
In this case the \numberratio{} varies from $0.7$ for $M_{\rm donor} = 25 M_{\odot}$ to $15$ for $M_{\rm donor} = 40 M_{\odot}$.
Compared to the standard models these subpopulations have significantly more BBHs mergers forming from BH--H-rich star systems with lower orbital periods between $P_{\rm orb} \approx 10$--$100~\mathrm{days}$ and $q<0.3$. 
Using the \citet{Claeys2014} prescription with \COSMIC{}, $\sim 77\%$ of BBH mergers form through CE evolution and the rest through stable MT. 

In the final row we vary the survivability of CE by now assuming the Pessimistic CE scenario. 
This variation had no significant effect on the resulting \numberratio{} for our \COSMIC{} BBH merger populations {\highlight at $0.1 Z_{\odot}$}. 
{\highlight This is because the majority of binaries affected by this variation involve massive donors that expand significantly during the Hertzsprung gap \citep[see also][Section 4.1]{Dominik2012}.
For our standard model in \COSMIC{}, we find that these affected donors tend to be more massive than those considered in our calculation.}

\begin{figure}
\centering
\includegraphics[width=0.5\textwidth]{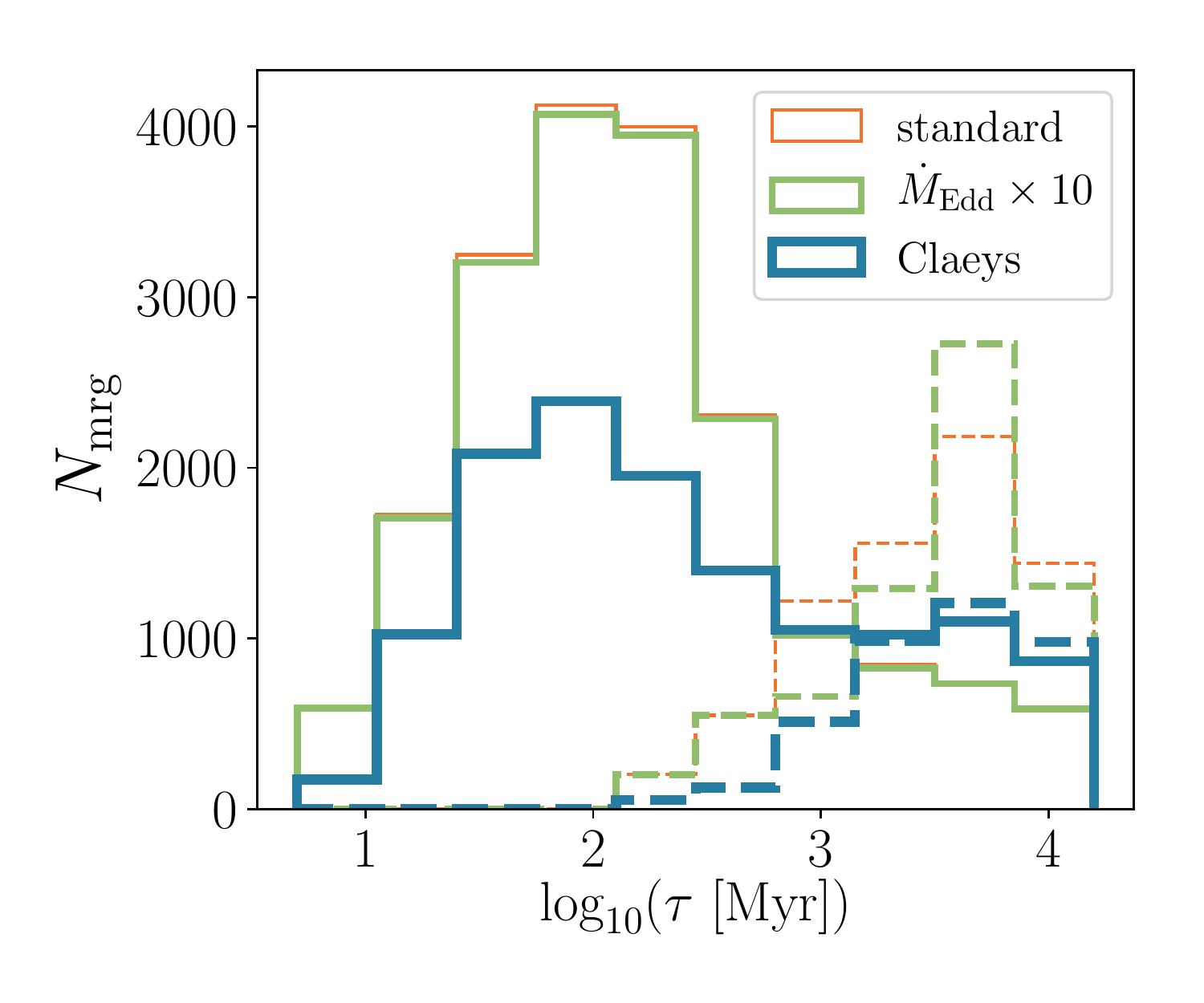}
\caption{ Histograms of $N_{\rm mrg}$ as a function of merger time $\tau$ for models ran with \COSMIC{} (solid lines) and models using detailed stellar and binary simulations of BH--MS systems (dashed lines). We show the standard model in orange (thin line), the model allowing for an accretion rate onto the BH of $10 \times \dot{M}_{\rm Edd}$ in green (medium line), and the model using $q_{\rm crit}$ criteria following \cite{Claeys2014} in \COSMIC{} in blue (thick line).  Here $\ensuremath{N_{\mathrm{mrg}}}$ includes all subpopulations per model. All models, except models with solar metallicity and $\alpha_{\rm CE}=5$, follow the same trend.
}
\label{fig:Ntau_histogram}
\end{figure}

The changes in modeling between \MESA{} simulations and \COSMIC{} not only influence the number of merging binaries, but also when they merge.
Figure~\ref{fig:Ntau_histogram} illustrates how \nmergercosmic{} (solid lines) and \nmergermesa{} (dashed lines) vary for different merger times for three representative models. 
In Fig.~\ref{fig:Ntau_histogram} we include all mass subpopulations per model as a proxy of the overall distribution across the mass range.
In all cases (except models with solar metallicity and $\alpha_{\rm CE}=5$) we find that \nmergercosmic{} is dominated by shorter merger times while \nmergermesa{} is dominated by longer merger times. 
These differences are due to the different dominate merger channels in each population. 
CE typically hardens the binary more efficiently than stable MT \citep{bavera2021}, which leads to shorter merger times, peaking at $\lesssim 100~\mathrm{Myr}$ \citep{Dominik2012,Eldridge2016}, for BBH mergers formed through CE. 
We see a difference in the trend in the $\alpha_{\rm CE}=5$ model because the post-CE separation in the \COSMIC{} systems is wider for $\alpha_{\rm CE}=5$ than in our standard model.
There are also a few cases of successful CE evolution among the \MESA{} results.
In solar metallicity models, the enhanced mass loss from stellar winds results in lower-mass BH, which leads to longer merger times and a flatter distribution.
The difference in merger times between CE evolution and stable MT will have an impact on the BBH merger rate across the history of the Universe.

\begin{figure}
\centering
\includegraphics[width=0.5\textwidth]{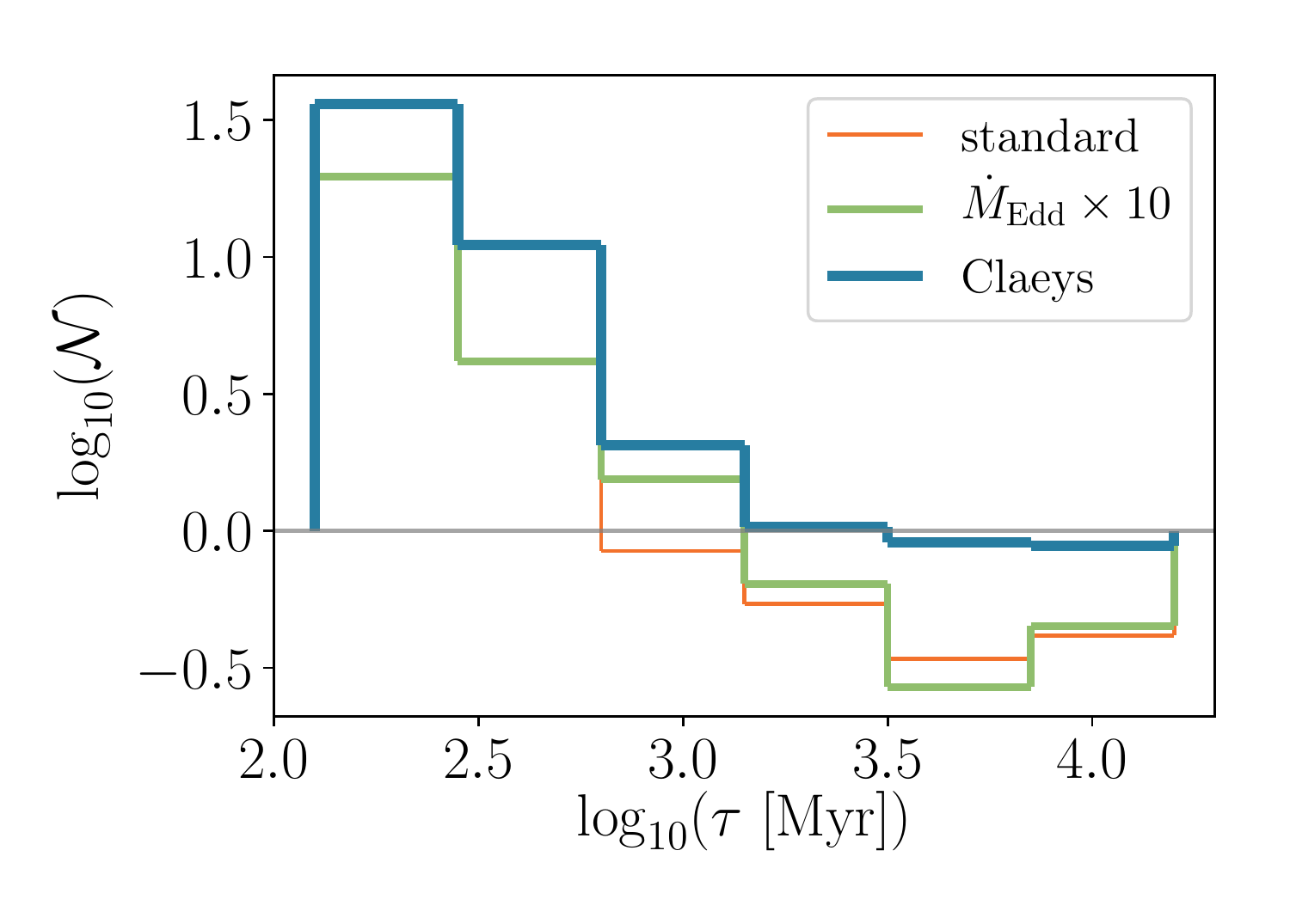}
\caption{ Same as Figure \ref{fig:Ntau_histogram} but here we show the \emph{number ratio} $\mathcal{N}$, the ratio of \nmergercosmic{} to \nmergermesa{}, binned as a function of merger time $\tau$.}

\label{fig:Ntau_ratio_hist}
\end{figure}

To further compare the differences in merger times, in Fig.~\ref{fig:Ntau_ratio_hist} we show the values of \numberratio{} $\mathcal{N} =\nmergercosmic{}/\nmergermesa{}$ binned as a function of merger times $\tau$. 
This shows that for BBH mergers with short merger times, the factor by which \COSMIC{} produces more BBH mergers than detailed simulations is much greater than when just considering the total as in Table~\ref{table:number_ratio_result}. 
However, while $\mathcal{N}$ for each subpopulation is generally $>1$, models using our standard $q_{\rm crit}$ criteria in \COSMIC{} produce fewer BBH mergers with long merger times.
{\highlight The merger time distribution will impact the overall merger rate and the mass distribution of merging BBHs at different redshifts.}
Measuring the merger rate as a function of redshift can potentially help to identify how BBHs form \citep{Fishbach2021,Santoliquido2021}, so it is important to have accurate predictions.

\begin{deluxetable}{ c c c c c}
\tablecaption{Same as Table \ref{table:number_ratio_result} but for \mathR{}, the ratio of our estimated subpopulation merger rate calculated for \COSMIC{} to the estimated merger rate calculated for \MESA{}. 
Since the total number of mergers is typically not more than an order of magnitude greater with \COSMIC{} than \MESA{}, a value $\gg 1$ corresponds to populations where \COSMIC{} is much more dominated by shorter merger times.   \label{table:tmrg_ratio} }
\tablehead{
\colhead{} & \multicolumn{4}{c}{Subpopulation} \\
\cline{2-5}
\colhead{Model} & $25 M_{\odot}$ & $30 M_{\odot}$ & $35 M_{\odot}$ &  $ 40 M_{\odot}$
}
\startdata 
{Standard} & 6 & 187 & 400& 145\\
{$\dot{M}_{\rm Edd} \times 10$} & 7.1 & 128 & 607 & 142\\
{$\alpha_{\rm CE} = 5$} & 0.24 & 168 & 554  & 293 \\ 
{$ Z_{\odot}$ } & - & - & - & 1.6 \\ 
Claeys & 5.7 & 80 & 422  & 744 \\
Pessimistic & 6 & 186 & 387 & 145 \\
\enddata
\end{deluxetable}

The differences in merger times shown in Fig.~\ref{fig:Ntau_histogram} and Fig.~\ref{fig:Ntau_ratio_hist} will have an impact on the resulting merger rate for BBHs. 
Table~\ref{table:tmrg_ratio} shows the values of \relativerate{} \mathR{} for all model variations and donor masses. 
Values of this ratio close to unity result from populations where BBH mergers modeled with \COSMIC{} and \MESA{} occur at similar rates. 
Large values result from populations where \COSMIC{} mergers are dominated by short merger times. 
The trends in this table follow roughly the same trends as in Table~\ref{table:number_ratio_result}. 
However, the variation between mass subpopulations and model variations highlights how the typical merger time varies. 
Compared to Table~\ref{table:number_ratio_result}, in some cases such as for $M_{\rm donor} = 40M_{\odot}$ with the standard model, $\mathcal{N}$ is close to unity, but this ratio \mathR{} is much larger than unity. 
These results show that not only do we get more mergers from \COSMIC{}, but the mergers are dominated by short times, which will yield a higher BBH merger rate soon after formation.

\section{Conclusions}

In this study we assessed how using detailed modeling of stellar physics, MT, and CE during the BH--H-rich star stage affects the final population of merging BBHs. 
We compared results from the rapid population synthesis code \COSMIC{} to detailed simulations using \MESA{}. 
For the models ran with \MESA{} we used a detailed method for MT and CE evolution \citep{Marchant2021}.
We find that modeling binary evolution with detailed simulations typically results in fewer mergers and longer merger times for BBHs. 

To investigate the impact of uncertainties in binary stellar evolution we varied metallicity, allowed for super-Eddington accretion, varied the CE efficiency between $\alpha_{\rm CE} = 1$ and $\alpha_{\rm CE} = 5$, and implemented two different prescriptions for critical mass ratios for CE stability in \COSMIC{}. 
For each model variation we identified and compared regions where successful BBH mergers occurred. 
We calculated a \numberratio{} \mathN{}, the ratio of the number of merging BBHs within a Hubble time between \COSMIC{} and \MESA{}. 
We also compared the distribution of merger times between the two populations and calculated a \relativerate{} \mathR{}, the ratio of the estimated merger rate for systems modeled with \COSMIC{} to those modeled with our detailed simulations using \MESA{}. 
Our main conclusions are:
\begin{enumerate}
    \item In all cases, models ran with \COSMIC{} and models ran with \MESA{} predict a different dominant formation channel for BBH mergers. 
    Merging BBHs modeled with \COSMIC{} are mainly formed via CE evolution.
    For models ran with \MESA{}, the dominant formation channel for merging BBHs is via stable MT. 

    \item  We find that many systems at high orbital periods, where \COSMIC{} assumes dynamical instability and predicts successful CE ejections, actually avoid unstable MT when modeled with our \MESA{} simulations. 

    \item For our models with $Z=0.1 Z_{\odot}$, we find fewer merging BBHs by up to a factor $15$, and a factor of $35$ for $\alpha_{\rm CE} = 5$, when using detailed stellar and binary physics with \MESA{} compared to our models simulated with \COSMIC{}. At solar metallicity \MESA{} models produce more merging BBHs than \COSMIC{}.
    
    \item Binaries modeled with \COSMIC{} are dominated by much shorter merger times compared to \MESA{} simulations. This, combined with the relative numbers results above, leads to significant differences in merger rates. \COSMIC{} models appear to overestimate the merger rates of BBHs by factors of $\sim\,5$--$500$.

\end{enumerate}
These results highlight how detailed modeling of stars in binary interactions can impact predictions for BBH populations and interpretation of GW sources. 

Consistent with our findings, other studies assessing the stability of massive stars to MT in various contexts have also found that stars are able to maintain dynamical stability in many more configurations than are commonly assumed in rapid population synthesis codes. 
Moreover, other population synthesis studies that have adopted stability criteria for MT based upon stellar-structure simulations have also found a more dominant role for stable MT in BBH formation \citep{neijssel_effect_2019,bavera2021,olejak2021impact,Shao2021}. 
Our results strengthen earlier conclusions and expand the implications of the dominance of stable MT: BBH systems are formed with wider orbits and hence systematically longer merger times, and as a result both the number of merging BBHs and the BBH merger rate can be significantly reduced.

A major current goal is to use GW observations, BBH masses, spins, and rates, in combination with binary stellar evolution predictions to assess the fraction of binaries formed through different channels \citep{Zevin2017,Bouffanais2019,Zevin2021}, and 
to constrain uncertain physical parameters that influence binary evolution, such as the CE efficiency \citep{wong2020,bavera2021,Zevin2021}, the mass-accretion efficiency \citep{bouffanais2020} or the distribution metallicities across the Universe \citep{bouffanais2021}. 
These constraints will become more precise as the GW source population grows \citep{Barrett2018}. 
However, constraints derived from models will be only as reliable and accurate as the models themselves.

Our results here, consistent with other studies, strongly motivate the need for population modeling that can account for stellar structure and evolution fully from ZAMS (not just in one particular binary phase and for a few specific mass slices, as we have done here) across the range of metallicities relevant to the formation of compact-object binaries. 
Parameter studies regarding initial conditions, SN kicks, and CE efficiency (despite the improved treatment used here) will still be needed, but fewer ad hoc assumptions will have to be incorporated compared to current rapid population synthesis analyses.

\begin{acknowledgements}

The authors thank Ying Qin, Aaron Dotter, Emmanouil Zapartas, and Jeff Andrews for their feedback and assistance with our \MESA{} simulations; Scott Coughlin for assistance with computational resources; Katie Brievik and Michael Zevin for help using \COSMIC{}; and Onno Pols and Rob Izzard for guidance of calculation of $\lambda$. 
We thank Tassos Fragos for insightful conversations, and thank the POSYDON collaboration for their support through the project (posydon.org). 
{\highlight We are grateful to the referee for constructive comments on the manuscript.}
M.G.-G.\ is grateful for the support from the Ford Foundation Predoctoral Fellowship. 
C.P.L.B.\ is supported by the CIERA Board of Visitors Research Professorship. 
P.M.\ acknowledges support from the FWO junior postdoctoral fellowship No.\ 12ZY520N. 
V.K.\ is supported by a CIFAR G+EU Senior Fellowship, by the Gordon and Betty Moore Foundation through grant GBMF8477, and by Northwestern University.
This work utilized the computing resources at CIERA provided by the Quest high performance computing facility at Northwestern University, which is jointly supported by the Office of the Provost, the Office for Research, and Northwestern University Information Technology, and used computing resources at CIERA funded by NSF PHY-1726951.
\end{acknowledgements}

\software{
Matplotlib \citep{Hunter2007}; 
NumPy \citep{vanderwalt2011};
Pandas \citep{mckinney-proc-scipy-2010}.
}

\bibliography{ms}
\bibliographystyle{aasjournal}

\end{document}